# Spontaneously generated coherence induced second-order susceptibility in atomic gases

Yueping Niu and Shangqing Gong

*State Key Laboratory of High Field Laser Physics, Shanghai Institute of Optics and Fine Mechanics, Chinese Academy of Sciences, Shanghai, 201800*

**Abstract:** The second-order susceptibility which vanishes in the electric-dipole approximation for an atom is induced by the spontaneously generated coherence. The spontaneously generated coherence considered in a lambda-type atomic system causes an indirect coupling between the lower two levels which acts equivalently as a DC-field and therefore makes possible the existence of second-order susceptibility.

The second-order susceptibility for an atom should vanish in the electric-dipole approximation because of symmetry [1, 2]. However, second harmonic generation (SHG) in atomic vapors has been observed by exerting laser pulses [3-9]. Clearly, in those experiments, the symmetry of the medium could have been broken during the laser excitation. Among all the mechanisms, the most plausible one is DC-field-induced SHG [4-8]. In these investigations, a DC-field was either applied or created by focused pump beam via multi-photon ionization processes. The coherent emission of a coherent excitation induced by two-photon resonance became allowed in the presence of the DC field. On the other hand, Ficek and Swain [10] have obtained the features of spontaneously generated coherence (SGC) from a Vee-type three level system without the need for parallel dipole moments. In this scheme, the ground state interacts with one of the upper two states by a laser field and the upper two states are coupled by a DC field. This system has been demonstrated to be equivalent to the usual three-level atomic system of Vee-type with parallel dipole moments. Subsequently, Joshi and Xiao [11] suggested realization of dark-state polaritons using SGC and also DC field without parallel dipole moments for substitution. Intrigued by the above projects, we here consider the generation of second-order susceptibility in a general discussed lambda-type three level system in the presence of SGC.

Consider a lambda-type system shown in Fig. 1, a resonant coupling field $\Omega_c$ drives the transition between levels |1> and |2> while a probe field $\Omega_p$ is applied to the transition |1> and |3>. $2\gamma_2$ and $2\gamma_3$ are the spontaneous decay rates of the excited state |1> to the ground states |2> and |3>.

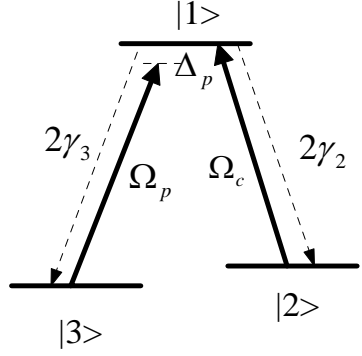

Fig. 1 Schematic energy-level diagram of the lambda-type system with the coupling field $\Omega_c$, the probe field $\Omega_p$ and decay rates $2\gamma_2$ and $2\gamma_3$.

When the two lower levels |2> and |3> are closely spaced such that the two transitions to the excited state interact with the same vacuum mode, spontaneously generated coherence could be present. Under the rotating wave approximation, the systematic density matrix in the interaction picture involving the SGC can be written as:

$$\dot{\rho}_{11} = -2(\gamma_2 + \gamma_3)\rho_{11} + i\Omega_p(\rho_{31} - \rho_{13}) + i\Omega_c(\rho_{21} - \rho_{12}),$$
$$\dot{\rho}_{33} = 2\gamma_3\rho_{11} + i\Omega_p(\rho_{13} - \rho_{31}),$$
$$\dot{\rho}_{23} = -i\Delta_p\rho_{23} + 2p\sqrt{\gamma_2\gamma_3}\rho_{11} + i\Omega_c\rho_{13} - i\Omega_p\rho_{21}, \quad (1)$$
$$\dot{\rho}_{13} = -(\gamma_2 + \gamma_3 + i\Delta_p)\rho_{13} - i\Omega_p(\rho_{11} - \rho_{33}) + i\Omega_c\rho_{23},$$
$$\dot{\rho}_{12} = -(\gamma_2 + \gamma_3)\rho_{12} + i\Omega_p\rho_{32} - i\Omega_c(\rho_{11} - \rho_{22}).$$

The above equations are constrained by $\rho_{11} + \rho_{22} + \rho_{33} = 1$ and $\rho_{ji}^* = \rho_{ij}$. $\Delta_p = \omega_{13} - \omega_p$ means the detuning of the probe field from the optical transition. The effect of SGC is very sensitive to the orientations of the atomic dipole moments $\vec{\mu}_{12}$ and $\vec{\mu}_{13}$. Here, the parameter $p$ denotes the alignment of the two dipole moments and is defined as $p = \vec{\mu}_{12} \cdot \vec{\mu}_{13} / |\vec{\mu}_{12} \cdot \vec{\mu}_{13}| = \cos\theta$ with $\theta$ being the angle between the two dipole moments. The term with $p\sqrt{\gamma_2\gamma_3}$ represents the quantum interference results from the cross coupling between spontaneous emission paths |1>-|2> and |1>-|3>. With the restriction that each field acts only on one transition, Rabi frequencies $\Omega_c$ and $\Omega_p$ are connected to the angle $\theta$ and represented by $\Omega_{c(p)} = \Omega_{c(p)}^0 \sin\theta = \Omega_{c(p)}^0 \sqrt{1-p^2}$. It should be stressed that only for small energy spacing between the two lower levels are the interference terms in Eq. (1) significant, otherwise the oscillatory terms will average out to zero and thereby SGC effect vanishes.

It is known that the response of the atomic medium to the probe field is govern by its polarization, $P = \varepsilon_0 (E_p \chi + E_p^* \chi^*)/2$, with $\chi$ being the susceptibility of the atomic medium. By performing a quantum average of the dipole moment over an ensemble of $N$ atoms, we find $P = N(\mu_{31}\rho_{13} + \mu_{13}\rho_{31})$. In order to derive the linear and nonlinear susceptibilities, we need to obtain the steady-state solution of the density matrix equations. In the present approach, an iterative method is used and the density matrix elements under the weak-probe approximation are expressed as:

$$\rho_{13}^{(1)} = \frac{i\Delta_p \Omega_p}{\Delta_p(\gamma_3 + \gamma_2 + i\Delta_p) - i\Omega_c^2}, \tag{2a}$$

$$\rho_{23}^{(2)} = \frac{p\sqrt{\gamma_2\gamma_3}(\gamma_2 + \gamma_3 + i\Delta_p)\Omega_p(\rho_{31}^{(1)} - \rho_{13}^{(1)})}{\gamma_3[\Delta_p(\gamma_2 + \gamma_3 + i\Delta_p) - i\Omega_c^2]}, \tag{2c}$$

$$\rho_{13}^{(2)} = \frac{i\Omega_c \rho_{23}^{(2)}}{(\gamma_2 + \gamma_3 + i\Delta_p)} = \frac{i\Omega_p \Omega_c p\sqrt{\gamma_2\gamma_3}(\rho_{31}^{(1)} - \rho_{13}^{(1)})}{\gamma_3[\Delta_p(\gamma_2 + \gamma_3 + i\Delta_p) - i\Omega_c^2]}. \tag{2d}$$

From the above equations, we can see that the occurrence of SGC gives rise to an extra coherence between the lower two levels |2> and |3> as long as the linear susceptibility exists. This term $\rho_{23}^{(2)}$ accordingly makes $\rho_{13}^{(2)}$ become nonzero. Therefore,

$$\chi^{(1)} = \frac{N\mu_{13}^2}{\varepsilon_0 \hbar} \frac{i\Delta_p}{-i\Omega_c^2 + \Delta_p(\gamma_2 + \gamma_3 + i\Delta_p)}, \tag{3}$$

$$\chi^{(2)} = \frac{N\mu_{13}^3}{\varepsilon_0 \hbar^2} \frac{2\Omega_c \Delta_p^2 p\gamma_2(\gamma_2 + \gamma_3)}{\sqrt{\gamma_2\gamma_3}(-i(\Omega_c^2 - \Delta_p^2) + \Delta_p(\gamma_2 + \gamma_3))^2 (i(\Omega_c^2 - \Delta_p^2) + \Delta_p(\gamma_2 + \gamma_3))}. \tag{4}$$

To present an intuitionistic impression, $\chi^{(2)}$ is displayed as a function of the probe detuning in Fig. 2 (solid curves). In the case of $p = 0$, that is no coherence is generated by spontaneous emission, although the linear susceptibility exists, the second-order susceptibility is always zero (Fig. 2 (a)). However, when SGC is taken into account, the second-order susceptibility appears, just as shown in Fig. 2 (b).

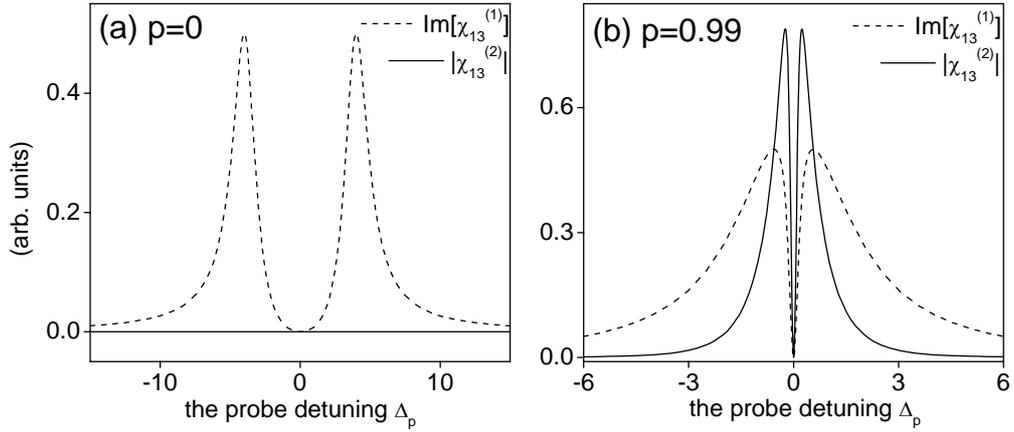

Fig. 2. The second-order susceptibility $|\chi_{13}^{(2)}|$ (solid curve) and linear absorption Im[ $\chi_{13}^{(1)}$ ] (dashed curve) as a function of the probe detuning $\Delta_p$. (a) without SGC, i.e., $p=0$; (b) with optimal SGC, i.e., $p=0.99$. Other parameters are $\Omega_c^0 = 4.0\gamma_2$, $\gamma_3 = \gamma_2$.

In the above system with SGC where the dipole moments are nonorthogonal, in addition to direct decay process, the spontaneous emission interference leads to an indirect coupling between the lower two states. This indirect interaction acts equivalently as a DC-field and therefore leads to the presence of second-order susceptibility which has been utilized for generation of the second-harmonic generation [7, 8]. Our investigation shows that this feature also occurs in the ladder and vee-type three level systems with SGC. Further discussions will be given elsewhere.

References:

[1] Y. R. Shen, *The Principles of Nonlinear Optics* (John Wiley & Sons, New York, 1984).

[2] T. F. Heinz and D. P. Divincenzo, Phys. Rev. A 43, 6249 (1990).

[3] S. Sullivan and E. L. Thomas, Opt. Commun. 25, 125 (1978).

[4] R. S. Finn and J. F. Ward, Phys. Rev. Lett. 26, 285 (1971).

[5] D. S. Bethune, Phys. Rev. A 23, 3139 (1981).

[6] L. Marmet, K. Hakuto, B. P. Stoicheff, Opt. Lett. 16, 261 (1991).

[7] K. Hakuta, L. Marmet and B. P. Stoicheff, Phys. Rev. A 45, 5152 (1992).

[8] K. Hakuta, L. Marmet and B. P. Stoicheff, Phys. Rev. Lett. 66, 596 (1991).

[9] C. Mullin, D. Kim, M. B. Feller and Y. R. Shen, Phys. Rev. Lett. 74, 2678 (1995).

[10] Z. Ficek, S. Swain, Phys. Rev. A 69, 023401 (2004).

[11] A. Joshi and M. Xiao, Eur. Phys. J. D 35, 547 (2005).